\documentclass[preprint,showpacs,amsmath,amssymb,prl,aps]{revtex4}

\usepackage{graphicx}
\usepackage{dcolumn}
\usepackage{bm}
\usepackage{amsmath}
\usepackage{amsfonts}
\usepackage{amssymb}

\begin{document}

\preprint{APL/rf-SET mixer}

\title{Mixing with the radiofrequency single-electron transistor}

\author{L. J. Swenson}
\author{D. R. Schmidt}
\author{J. S. Aldridge}
\author{D. K. Wood}
\author{A. N. Cleland\footnote{Electronic mail : cleland@physics.ucsb.edu}}
\affiliation{Department of Physics, University of California at
Santa Barbara, Santa Barbara, CA 93106}

\date{\today}

\begin{abstract}
By configuring a radio-frequency single-electron transistor as a mixer, we demonstrate a unique
implementation of this device, that achieves good charge sensitivity with large bandwidth about a
tunable center frequency. In our implementation we achieve a measurement bandwidth of 16 MHz, with
a tunable center frequency from 0 to 1.2 GHz, demonstrated with the transistor operating at 300 mK.
Ultimately this device is limited in center frequency by the $RC$ time of the transistor's center
island, which for our device is $\sim 1.6$ GHz, close to the measured value. The measurement
bandwidth is determined by the quality factor of the readout tank circuit.
\end{abstract}


\maketitle

In mesoscopic experiments requiring ultra-sensitive charge
detection, the single electron transistor (SET) has become widely
recognized as one of the most suitable charge amplifiers. With the
first experimental realization by Fulton and Dolan
\cite{fulton:109}, and theoretical description by Averin and
Likharev \cite{averin:345}, the SET is a small-capacitance
metallic island onto and off of which electrons can tunnel via
source and drain leads. At low temperatures $k_B T \ll
e^2/2C_{\Sigma}$, where $C_{\Sigma}$ is the island capacitance,
tunneling is suppressed for source-drain voltages $V$ in the range
$|V| < e/2C_{\Sigma}$, known as the Coulomb blockade. By
introducing a gate lead, capacitively coupled to the metallic
island, the electrostatic energy of the island, and hence the
tunnelling rate, can be manipulated by voltages on the gate. The
drain-source current is extremely sensitive to the gate charge,
yielding a very low noise charge-to-current transducer. Successful
operation of the SET requires that the tunnel resistance $R_T$ of
the drain and source junctions satisfy $R_T \geq R_{K} = h/e^2
\approx$  25.8 k$\Omega$, the quantum of resistance.

Despite its clear advantages, the SET has until recently suffered
from a major drawback: The large tunnel resistance $R_T$, coupled
with the unavoidable stray capacitance of the wiring,
$C_{\mathrm{stray}} \sim 10^{-12}$ F, limits the output bandwidth
to at best $1/2 \pi R_T C_{\mathrm{stray}} \sim$ 1-10 MHz. High
frequency signals, necessary for measurement of the dynamics of
systems such as a nanomechanical resonator\cite{knobel:291} or an
excited Cooper-pair box\cite{lehnert:027002}, remain undetected
when operating with the standard SET configuration. Two recent
innovations have demonstrated significant improvements on this
limiting behavior, greatly increasing the SET's spectral range.
The first approach was to use a series inductance, resonating with
the stray lead capacitance, to create a tank circuit that roughly
impedance matches the SET to a low-impedance cable. This
innovation is termed the radio-frequency single-electron
transistor, or rf-SET \cite{schoelkopf:1238}, and has been used to
achieve measurement bandwidths in excess of 100 MHz, with
intrinsic charge noise below $10^{-5}~e / \sqrt{\mathrm{Hz}}$
\cite{devoret:1039}. The second approach was to use the non-linear
response of the SET current to the gate signal to implement the
SET as a radiofrequency mixer, still limited by the $RC$ charging
time to a narrow bandwidth, but allowing measurements at center
frequencies tunable up to $1-10$ GHz, limited by the intrinsic
$1/RC_{\Sigma}$ bandwidth of the SET island \cite{knobel:532}.

\begin{figure}
\includegraphics[width=0.7\linewidth]{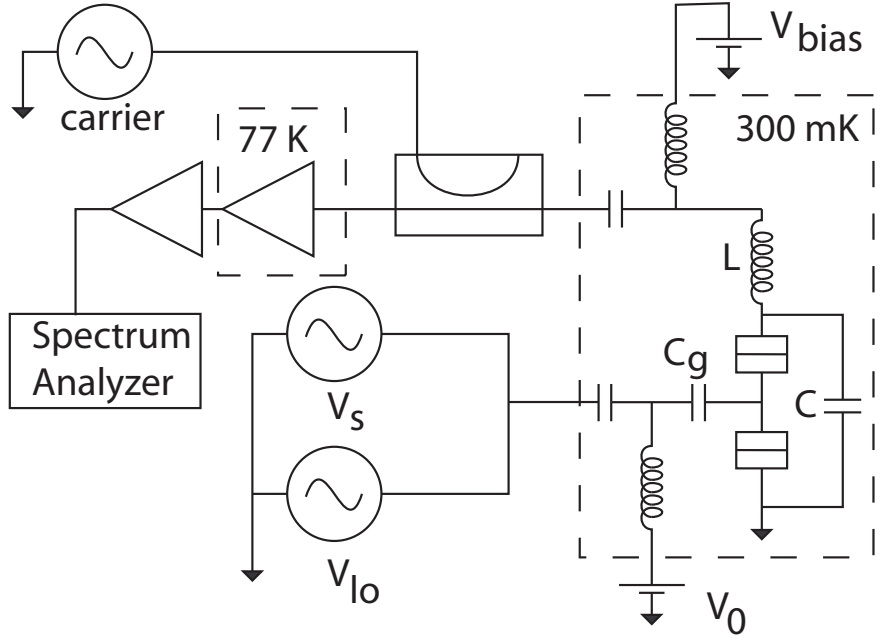}
\caption{Schematic of the measurement circuit. Dashed outline
shows cryogenic part of experiment, and $L$ and $C$ indicate the
tank circuit elements; $C_g$ is the SET gate capacitance. $V_s$
and $V_{lo}$ are rf signal amplitudes; $V_0$ is the dc gate bias.
The reflected power preamplifier is cooled to 77 K.}
\label{fig:circuit}
\end{figure}

Here we demonstrate experimentally that the large bandwidth of the
rf-SET, and the tunability of the SET mixer, can be simultaneously
achieved. A schematic of the measurement setup is shown in Fig.\
\ref{fig:circuit}.  An all-aluminum SET, with a total tunnelling
resistance $R_T \approx 95~\mathrm{k}\Omega$, island capacitance
$C_{\Sigma} \approx 550$ aF,  and input gate capacitance $C_{g}
\approx$ 22 aF, was fabricated by standard shadow evaporation on a
semi-insulating GaAs chip. The chip was glued to a printed circuit
board mounted in a metal box, and mounted on the cold stage of a
300 mK $^3$He cryostat. The drain of the SET was grounded and the
source lead was connected in series with a 390 nH inductor, in a
standard rf-SET configuration; the resonant capacitance is mostly
that of the on-chip drain lead to the SET. The other end of the
inductor was connected to a 50 $\Omega$ semi-rigid coaxial cable,
interrupted by a bias tee for applying dc bias power, before
passing out of the cryostat to room-temperature electronics. This
configuration gave a resonant tank circuit frequency $f_{LC} =
326$ MHz.  A second 50 $\Omega$ coaxial cable was connected to the
gate of the SET, interrupted by a bias tee to allow dc voltages to
tune the operating point of the SET.

\begin{figure}
\includegraphics[width=\linewidth]{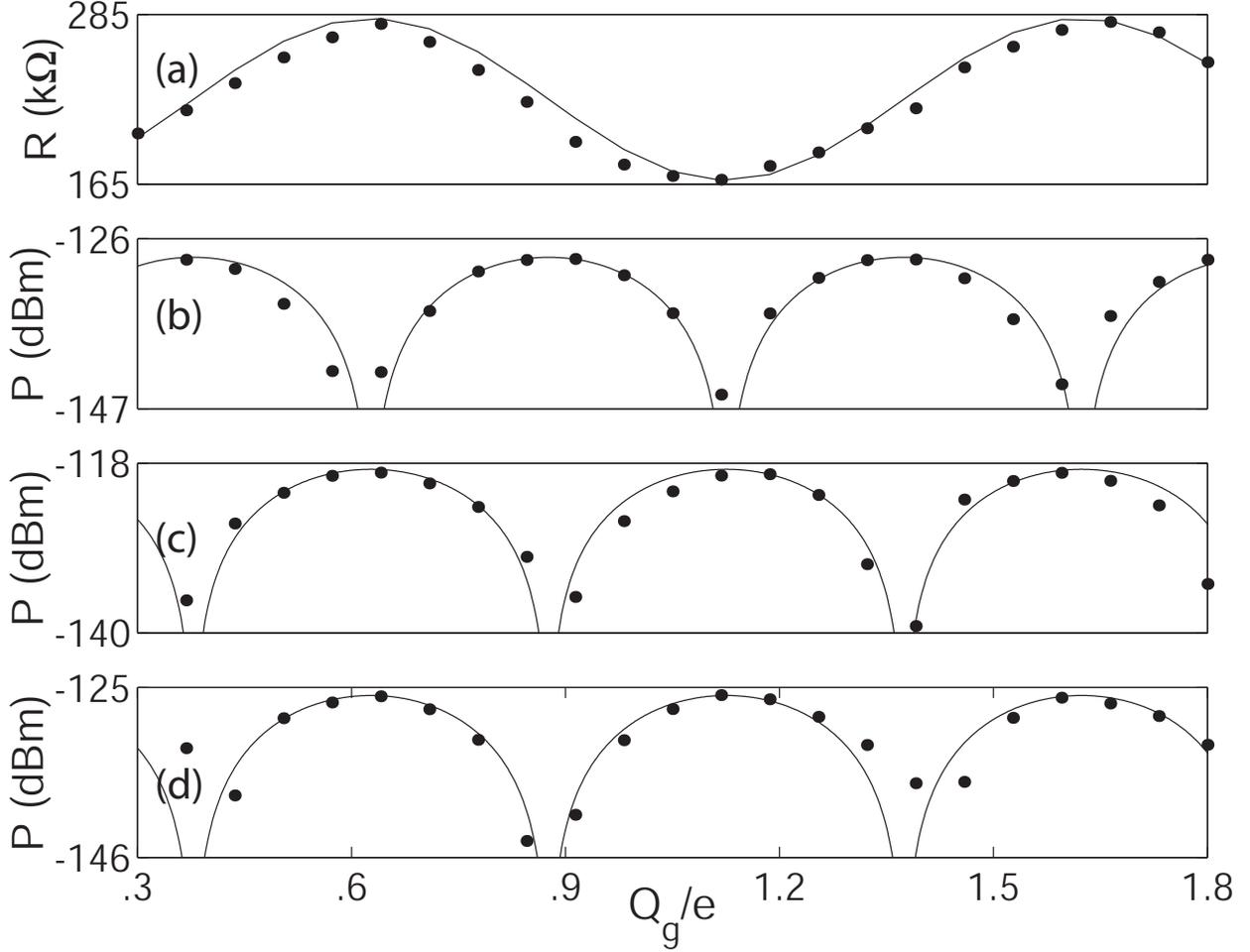}
\caption{SET response as a function of dc gate charge: (a)
Source-drain differential resistance $R = dV/dI$; solid line is a
sinusoidal fit. (b) First harmonic reflected power for with a 0.5
MHz, 0.01$e$ amplitude rf signal applied to the gate (small
amplitude limit). (c) Second harmonic reflected power, with a 0.5
MHz, 0.09$e$ large-amplitude rf signal on the gate. (d) Reflected
power of heterodyne mixer at the sideband $|f_{lo}-f_{s}|$, where
$f_{lo} = 412$ MHz and $f_{s} = 410$ MHz. $C_g V_{lo}$ and $C_g
V_{s}$ were held fixed at 0.19$e$ and 0.15$e$, respectively. }
\label{fig:figure2}
\end{figure}

By applying a dc bias to the drain, and measuring the power
reflected at the carrier frequency $f_c$ from the tank circuit,
set to the tank circuit resonance frequency $f_c = f_{LC}$, the
device was first operated as an rf-SET in the superconducting
state\cite{fulton:1307}. In this implementation, a dc gate voltage
$V_0$, and a rf gate signal $V_s$ at frequency $f_s$, modulates
the SET differential drain-source resistance, and thus modulates
the reflected power amplitude.  As the drain-source resistance
modulation $\Delta R$ was always very small, the reflection
coefficient is to a good approximation linearly dependent on
$\Delta R$. The resistance modulation is a function of the gate
voltage, $V_g = V_0+V_s \cos (2 \pi f_s t)$. Fig.\
\ref{fig:figure2}(a) shows the measured dependence of $\Delta R$
on the gate charge $Q_g = C_g V_0$, with no applied rf signal. We
operated at a bias point, near the Josephson quasi-particle peak,
where the SET source-drain differential resistance change $\Delta
R = dV/dI(Q_g)-dV/dI(0)$ has a sinusoidal dependence on the gate
charge,
\begin{equation}
  \Delta R \approx R_{g} \cos \left ( \frac{2 \pi Q_g}{e} \right ),
 \label{eqn01}
\end{equation}
where $R_{g}$ is the range accessible by gating. A fit of Eq.\
(\ref{eqn01}) to the data is displayed in Fig.\
\ref{fig:figure2}(a). Using this approximation, the time-dependent
resistance change of the SET, with an rf signal at the gate, is
given by
\begin{equation}
  \Delta R(t) \approx R_{g} \cos \left ( \frac{2 \pi C_{g}}{e} \left [ V_0+V_s \cos(2 \pi f_s t)
  \right ] \right ).
 \label{eqn02}
\end{equation}
For small rf amplitudes $C_g V_s \ll e$, with the gate dc bias
$V_0$ chosen so that Eq.\ (\ref{eqn02}) has the steepest slope, at
e.g. $C_g V_0 = -e/4$, the SET resistance change is linear in the
gate amplitude. For these small resistance changes, the spectrum
of the power reflected from the tank circuit contains two
sidebands at $f_c \pm f_s$, whose amplitudes are also linearly
dependent on the rf gate amplitude $C_g V_s$; this is the usual
mode of operation for the rf-SET.

For larger gate amplitudes, the SET resistance becomes non-linearly related to the gate signal, and
the reflected power spectrum then contains additional sidebands at $f_c \pm n f_s$, where $n$ takes
on positive integer values. The strength of the harmonic terms depends nonlinearly on the gate
signal amplitude, as well as on the gate dc bias point.

In Figs.\ \ref{fig:figure2}(b) and \ref{fig:figure2}(c) we show
the measured sideband reflected power as a function of the gate
charge $C_g V_0$ for the $n=1$ and $n=2$ harmonics, with signal
amplitude $V_s$ fixed at $0.01 e$ and $0.09 e$, respectively. As
can be seen by comparing to Fig.\ \ref{fig:figure2}(a), the $n=1$
and, in general, the odd harmonics give maximum amplitude when the
dc gate charge is adjusted to put Eq.\ (\ref{eqn01}) at a point of
steepest slope, while the $n=2$ and even harmonics are maximized
when the gate charge puts Eq.\ (\ref{eqn01}) at an extremum.

\begin{figure}
\includegraphics[width=\linewidth]{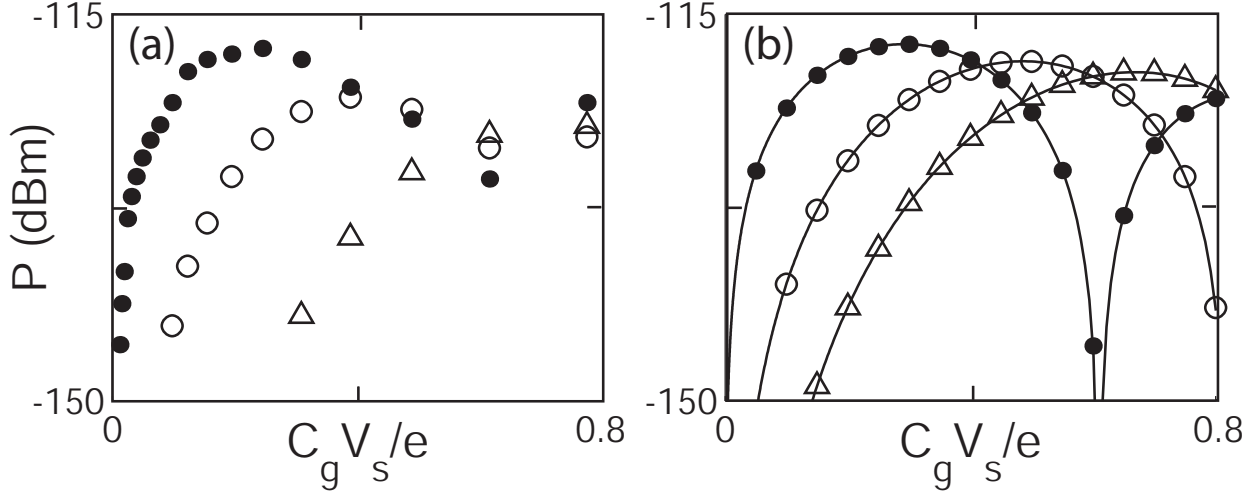}
\caption{Reflected power as a function of rf gate amplitude $C_g
V_s$: (a) Measured sideband reflected power for the first
($\bullet$), second ($\circ$), and third ($\triangle$) harmonics,
as a function of $C_g V_s$ ranging from 0 to $0.8 e$. The gate rf
signal was at 4 MHz. (b) Model dependence for the
first($\bullet$), second($\circ$), and third ($\triangle$)
sideband reflected power, under the same conditions as (a). }
\label{fig:figure3}
\end{figure}

In Fig.\ \ref{fig:figure3}(a) we show the dependence of the
reflected power for the first three harmonics on the rf gate
amplitude $V_s$. As noted above, the odd harmonics were measured
at a different dc gate charge than the even harmonics. To model
this dependence, we Fourier transform Eq.\ (\ref{eqn02}) for each
harmonic. The Fourier coefficients for $n \leq 3$ can be expressed
by
\begin{equation}
  c_{n} = R_{g} J_{n} \left ( \frac{2\pi C_g V_s}{e} \right ),
\label{eqnA}
\end{equation}
where $J_{n}$ is the $n^{\mathrm{th}}$ order Bessel function. In Fig.\ \ref{fig:figure3}(b) we show
the model rf-SET response for the first three harmonics, equal to $20\log_{10}(c_{n})$ plus an
offset that depends on the overall reflection coefficient and incoming power level. Agreement is
relatively good.

The response we have described so far represents a form of homodyne detection with the rf-SET.
However, the nonlinear response of the SET differential resistance to the gate charge allows the
rf-SET to also be used as a heterodyne mixer, sensitive to \emph{small} signal variations on the
gate. This is achieved by applying a large-amplitude local oscillator (LO) to the gate at frequency
$f_{lo}$, in addition to the small signal to be detected at frequency $f_s$. With both signals
applied to the SET gate, the SET resistance response can be modelled using the approximate response
given by Eq.\ (\ref{eqn01}),
\begin{equation}
  \Delta R \approx R_{g} \cos \left ( \frac{2 \pi C_{g}}{e} \left [ V_0 + V_{lo} \cos(2 \pi f_{lo} t) +
  V_s \cos(2 \pi f_s t) \right ] \right ),
\label{eqn03}
\end{equation}
where $V_{lo}$ and $V_s$ are the local oscillator and signal amplitudes, respectively.

In the nonlinear response regime of the SET, the resistance change given by Eq. (\ref{eqn03})
contains frequency components $f_{mn}$, where $f_{mn} = |m f_{lo} \pm n f_s|$, for positive
integers $m$ and $n$.  For either $m$ or $n$ equal to zero, this is the previously-described
homodyne detection. The spectrum of the power reflected from the SET tank circuit then has
components $f_c \pm f_{mn}$ about the carrier frequency $f_c$. A Fourier representation of Eq.\
(\ref{eqn03}) thus contains an infinite number of terms with non-zero coefficients; we only
consider those within the output bandwidth of the rf-SET. For small rf signal amplitude $V_s$, the
Fourier coefficients will be most appreciable for $m,n \leq 2$. When $m+n$ is even, the Fourier
coefficient is a maximum when the resistance change $\Delta R$ given by Eq.\ (\ref{eqn01}) is at an
extremum; when $m+n$ is odd, the coefficient is greatest when the derivative of $\Delta R$ with
respect to the gate charge is maximized. The case for $|f_{lo}-f_{s}|$ ($m = n = 1$, $m+n$ even) is
shown in Fig.\ \ref{fig:figure2}(d).

\begin{figure}
\includegraphics[width=\linewidth]{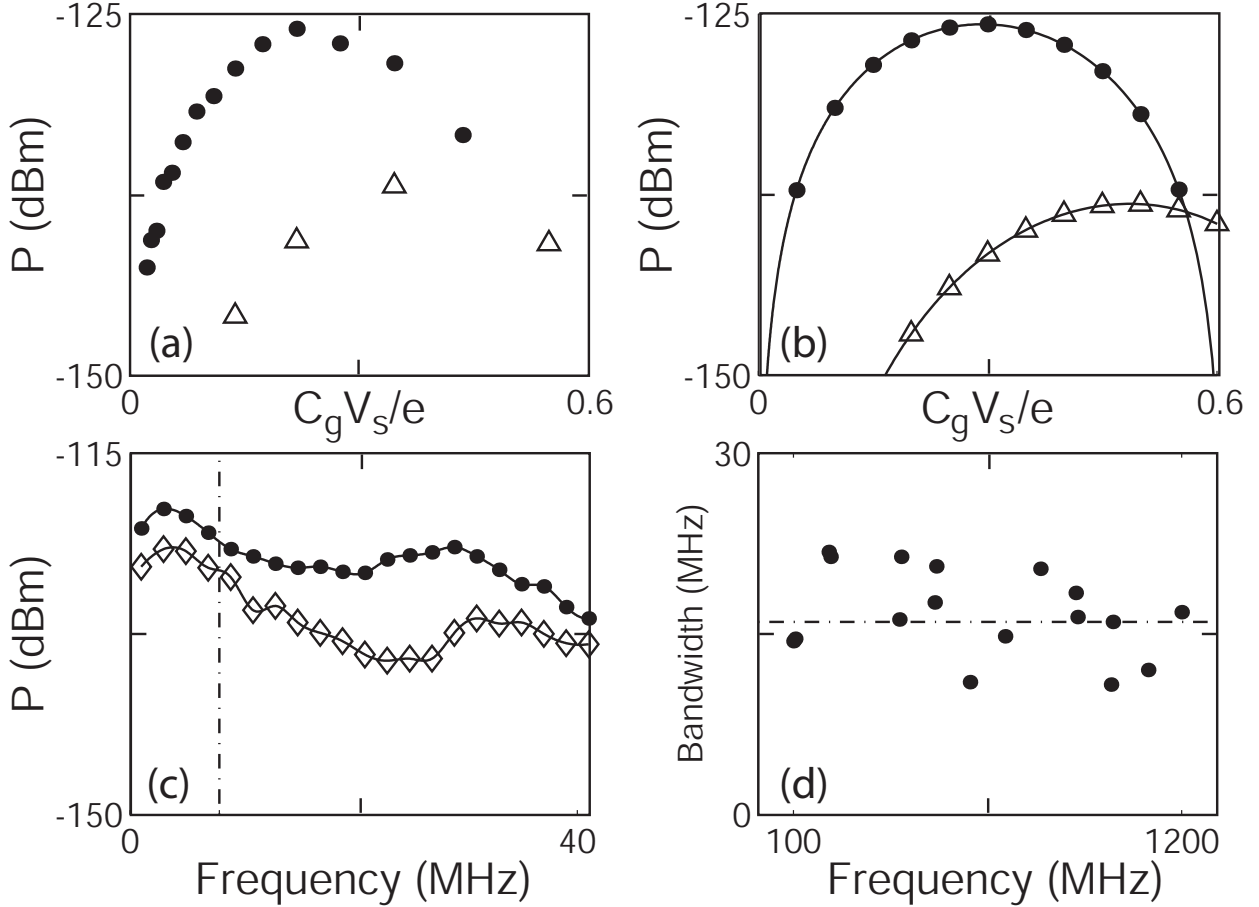}
\caption{Characterization of the rf-SET heterodyne response, as a
function of gate charge and frequency. (a) Measured sideband
reflected power for sidebands $|f_{lo}-f_{s}|~(\bullet)$ and
$|2f_{lo}-2f_{s}|~(\triangle)$ as a function of $C_g V_s$, which
ranges from 0 to $0.6 e$. $C_g V_{lo}$ was held fixed at $0.17 e$,
the dc gate bias was set optimally, $f_{lo} = 410$ MHz, and $f_s =
412$ MHz. (b) Modelled sideband reflected power under the same
amplitude conditions as in (a). (c) First harmonic reflected power
as a function of frequency separation $\Delta f = f_c-f_s$ for
homodyne detection ($\bullet$), and when operated as a mixer, with
$\Delta f = f_c-|f_{lo}-f_{s}|$, where $f_{lo} = 400$ MHz
($\Diamond$). The dashed line is 3 dB down from the maximum
reflected power, yielding a 16 MHz bandwidth. (d) Measured mixer
bandwidth as a function of $f_{lo}$. The dashed line is at 16
MHz.} \label{fig:figure4}
\end{figure}

Using our model, under optimal dc bias conditions the Fourier coefficients for $m,n \leq 2$ can be
expressed as
\begin{equation}
  c_{mn} = R_{g} J_{m} \left ( \frac{2 \pi C_g V_{lo}}{e} \right ) \times
  J_{n} \left ( \frac{2 \pi C_g V_s}{e} \right )
\label{eqn04}
\end{equation}
In  Fig.\ \ref{fig:figure4}(a) we show the measured dependence of the reflected power sideband
components $m = n = 1$ and $m = n = 2$ on $C_g V_s$, with dc gate bias optimized separately for
each component. In Fig.\ \ref{fig:figure4}(b) we show the model response, equal to $20
\log_{10}(c_{mn})$ plus an offset, in fairly good correspondence with Fig.\ \ref{fig:figure4}(a).
In Fig.\ \ref{fig:figure4} (c) we show the dependence of the reflected power on frequency
separation $f_c-f_s$, for both homodyne and heterodyne detection.  In Fig.\ \ref{fig:figure4}(d)
the mixer bandwidth as a function of $f_{lo}$ is displayed.

For both homodyne and heterodyne detection, the first harmonic ($n = 1$) yields the greatest charge
sensitivity for small signals $V_s$. When operated as a mixer, the charge sensitivity is maximized
when $m = 1$ (first LO component) and $C_g V_{lo} \sim 0.293 e$ (maximum of the Bessel function
corresponding to the LO signal). However, due to the form of $c_{mn}$, the signal when mixing will
be reduced by about 5 dB compared to homodyne detection.

We have measured the charge sensitivity of this device, operating
both as an rf-SET and as a heterodyne mixer. The calculation of
charge sensitivity for the rf-SET has been previously published
\cite{aassime:4031, roschier:1274}. The charge noise for optimal
bias conditions was measured to be $\delta q_s \leq 2 \times
10^{-3} e / \sqrt{\mathrm{Hz}}$ in rf-SET mode, and increased
slightly to $\delta q_s \leq 5 \times 10^{-3} e /
\sqrt{\mathrm{Hz}}$ when operated as a mixer. However, these
values were dominated by the noise in the 77 K preamplifier, used
to amplify the reflected power from the tank circuit; a 4.2 K
mounted preamplifier would likely yield better noise figures
\cite{schoelkopf:1238}. These indicate that the performance of the
rf-SET when used as a mixer is not significantly worse than when
used in homodyne detection.

In conclusion, we have demonstrated mixing with an rf-SET, allowing tuning of the 16 MHz
measurement bandwidth around a center frequency which could be set up to 1.2 GHz. The center
frequency is ultimately limited by the $RC$ time constant of the SET center island, here estimated
to give a limit of $1/2 \pi R C \sim $ 1.6 GHz. This technology will facilitate transmission-style
measurements requiring both small signal detection and sensitivity over a broad range of signal
frequencies and amplitudes.

\begin{acknowledgments}
We thank Bob Hill for processing support, and we acknowledge financial support provided by the NASA
Office of Space Science under grants NAG5-11426.
\end{acknowledgments}

\bibliographystyle{aip}

\bibliography{papers}

\begin{thebibliography}{10}

\bibitem{fulton:109}
T.~A. Fulton and G.~J. Dolan,
\newblock Physical Review Letters {\bf 59}, 109 (1987).

\bibitem{averin:345}
D.~V. Averin and K.~K. Likharev,
\newblock J. Low Temp. Phys. {\bf 62}, 345 (1986).

\bibitem{knobel:291}
R.~G. Knobel and A.~Cleland,
\newblock Nature {\bf 424}, 291 (2003).

\bibitem{lehnert:027002}
K.~W. Lehnert et~al.,
\newblock Physical Review Letters {\bf 90}, 027002 (2003).

\bibitem{schoelkopf:1238}
R.~J. Schoelkopf, P.~Wahlgren, A.~A. Kozhevnikov, P.~Delsing, and D.~E. Prober,
\newblock Science {\bf 280}, 1238 (1998).

\bibitem{devoret:1039}
M.~H. Devoret and R.~J. Schoelkopf,
\newblock Nature {\bf 406}, 1039 (2000).

\bibitem{knobel:532}
R.~Knobel, C.~S. Yung, and A.~N. Cleland,
\newblock Applied Physics Letters {\bf 81}, 532 (2002).

\bibitem{fulton:1307}
T.~A. Fulton, P.~L. Gammel, D.~J. Bishop, L.~N. Dunkleberger, and G.~J. Dolan,
\newblock Physical Review Letters {\bf 63}, 1307 (1989).

\bibitem{aassime:4031}
A.~Aassime, D.~Gunnarsson, K.~Bladh, P.~Delsing, and R.~Schoelkopf,
\newblock Applied Physics Letters {\bf 79}, 4031 (2001).

\bibitem{roschier:1274}
L.~Roschier et~al.,
\newblock Journal of Applied Physics {\bf 95}, 1274 (2004).

\end{thebibliography}
\end{document}